\documentstyle[psfrag,aps,preprint,epsfig,axodraw]{revtex}

\begin{document}

\tighten

\preprint{TU-697}
\title{Raising Sfermion Masses by Adding Extra Matter Fields}
\author{Motoi Endo, Masahiro Yamaguchi and Akira Yotsuyanagi}
\address{ Department of Physics, Tohoku University,
Sendai 980-8578, Japan}
\date{October 2003}
\maketitle
\begin{abstract} 
The renormalization group flow of soft supersymmetry breaking masses is
sensitive to the field contents of the theory one considers. We point out 
that the addition of extra vector-like matter fields to
the minimal supersymmetric standard model raises the
masses of squarks and sleptons relative to those of gauginos. 
We discuss its phenomenological implications. Besides an obvious effect to
the superparticle mass spectrum, we find that radiative corrections from
heavier stop loops  increase the lightest CP-even Higgs boson mass. 
We also discuss impact on models with no-scale boundary conditions. It turns
out that, unlike the minimal case, staus can become heavier than a B-ino like
neutralino, which is cosmologically favored.
\end{abstract} 

\clearpage

It has  widely been believed that low energy supersymmetry 
(SUSY) \cite{Weinberg} 
is the most
promising approach to solve the naturalness problem on the electroweak
scale inherent in the standard model of particle physics. If this line
of reasoning is correct, one of the most interesting tasks is to
reveal the nature of the mechanism of supersymmetry breaking and its
mediation to the standard model sector.  Superparticle masses
which are evaluated at the electroweak scale are supposed to be
given at high energy scale in a hypothetical fundamental theory.  A
key ingredient to connect the quantities at the different scales is
renormalization group (RG) evolution.
 
It is known that the RG flow depends on the field contents of the
theory one considers. In this paper, we will examine how the addition
of extra vector-like matter fields to the minimal supersymmetric
standard model (MSSM) affects the RG evolution and superparticle mass
spectrum. 
Though the MSSM particle contents successfully explain the
unification of three gauge coupling constants, there is still some
room to add extra vector-like matter fields with the restriction that
they should constitute full $SU(5)$ multiplets in order not to destroy
the success at least  at one loop level.  Introduction of extra matter fields
is often considered to solve some difficulty of particle physics models. 
Examples include  a hadronic axion model \cite{hadronic-axion}, an
attractive solution to the strong CP problem.

The key observation we make in this paper is that, as we will discuss
shortly,  the change of the
RG evolution of the gauge coupling constants as well as the gaugino
masses due to the presence of the extra matter fields raises the sfermion 
masses  when compared to
the gaugino masses at the electroweak scale. This observation has a
lot of phenomenological implications. Besides an obvious remark on the
modification of the superparticle mass spectrum which will hopefully
be measured in future collider experiments \cite{Tsukamoto-etal}, we
will point out that the larger stop mass implied by the heavier
sfermion mass spectrum makes the Higgs boson mass large due to
radiation corrections from top-stop loop \cite{HiggsMass}. In fact, the
experimental Higgs mass bound gives a rather severe constraint on
models where scalar masses given at high energy scale are small. The
RG effect with the extra matter fields will somewhat relax the
constraint.

We will also discuss implications to models with no-scale boundary
conditions. In this scenario, the scalar masses vanish at the
boundary, and thus it can be a natural solution to the supersymmetric
flavor problem. 
This type of boundary conditions was realized
originally in the no-scale model \cite{no-scale-model} and also in the
context of heterotic string theory \cite{Witten}. It was recognized
\cite{IKYY} that the vanishing scalar mass as well as a vanishing
$A$-parameter is a common feature of the models where the hidden and
observable sectors are appropriately separated in the K\"ahler
potential, and then the gaugino masses can be  the only source of the
SUSY breaking masses. It is interesting to note that the splitting may
naturally be realized in the geometrical setting where the hidden-sector
brane is sequestered from the standard-model brane
\cite{Randall-Sundrum,Luty-Sundrum}. In this setup, the gauginos can
acquire SUSY breaking masses if they propagate in the bulk and couple
to the SUSY breaking fields on the hidden brane
\cite{gaugino-mediation}.
 Despite this attractive feature, the minimal
setup with the RG evolution starting from the grand-unified-theory
(GUT) scale faces a serious phenomenological difficulty. The point is
a coincidental degeneracy in masses of right-handed sleptons and B-ino.
A previous study showed that the cosmological requirement that the
neutralino to be the lightest superparticle (LSP) (not charged
slepton) gives very stringent constraint on the upperbound of the
superparticle mass spectrum \cite{IKYY,KLNPY,Drees-Nojiri}. Moreover with this
constraint, the predicted Higgs mass would be lower than its
experimental bound, and thus this interesting idea would conflict with
experiments. There have been proposed several mechanisms to avoid this
problem in the literature, including possible RG flow above the GUT
scale~\cite{KMY,PolonskyPomarol,SchmaltzSkiba}, light gravitino or 
axino LSP scenario,  
and also non-universal gaugino masses
\cite{KomineYama}.  Here we will propose an alternative solution with
the addition of the extra matter fields.

\

We begin by discussing  the RG evolution of soft
SUSY breaking mass parameters. The RG equations (RGEs) for the
gauge coupling constants $\alpha_i$ in $N=1$ supersymmetry are written
\begin{eqnarray}
  \mu \frac{d \alpha}{d \mu} &=& - \frac{b}{2\pi} \alpha^2 \ ,\\
   b&=& C_2(G)-\sum_{\mbox{chiral}} T(R) \label{eq:bi}
\end{eqnarray}
where $C_2(G)=N$ for $G=SU(N)$ and $T(R)$ is defined as 
$\mbox{Tr} T^a T^b=T(R)\delta^{ab}$ with $T(\mbox{fund})=1/2$ and 
in the second term of (\ref{eq:bi}) summation over the chiral multiplets 
is understood.  
When one considers MSSM particle contents with
additional extra vector-like matter fields, the above become
\begin{eqnarray}
  \mu \frac{d \alpha_i}{d \mu} = - \frac{\beta_i - N_{\mbox{ex}}}{2\pi} 
  \alpha_i^2 \ ,
\end{eqnarray}
where $i$ runs from 1 to 3, with $\beta_i=(3, -1, -33/5)$ for $SU(3)_C
\times SU(2)_L \times U(1)_Y$.  Here we have assumed that the extra
matter multiplets consist of full multiplets in terms of $SU(5)$ GUT, 
which is requisite not to destroy the successful
gauge coupling unification in the MSSM, and we denote the number of
the extra vector-like matter multiplets by $N_{\mbox{ex}}$, whose
normalization is given such that $N_{\mbox{ex}}=1$ for one pair of
${\bf 5}$ and ${\bf \bar 5}$ representations and $N_{\mbox{ex}}=3$ for
one pair of ${\bf 10}$ and ${\bf \overline{10}}$ representations.

Similarly the RGEs for the gaugino masses are given as
\begin{eqnarray}
  \mu \frac{d M_i}{d \mu} = - \frac{\beta_i - N_{\mbox{ex}}}{2\pi}
   \alpha_i M_i \ .
\end{eqnarray}

Throughout this paper, we consider the case where  the extra matter fields 
do not have large Yukawa coupling to the ordinary quarks and leptons and 
to the Higgs fields. 
In this case the RG evolution of the SUSY breaking scalar masses
is not modified by the extra matter fields at one-loop level.
  
For simplicity, we assume that all extra matter fields have a common
mass, $M_{\mbox{ex}}$. We also assume that they do not mediate any
non-trivial SUSY breaking (unlike gauge mediation) and that the
threshold effects to soft masses, when they decouple, are negligibly
small, which is justified when the soft SUSY breaking $B$-parameters
for the extra matter fields are not much larger than the gaugino
masses and the number of the extra matter fields is not extremely
large. The latter condition is always  fulfilled in our case, because
it is restricted by the perturbativity of the gauge coupling
constants. With this setup, we simply solve the RGEs with the extra
multiplets above $M_{\mbox{ex}}$, and below this scale use is 
made of the RGEs of the MSSM, with  the trivial matching condition 
without any threshold corrections imposed at $M_{\mbox{ex}}$.

\begin{figure}[htbp]
   \begin{center}
     \includegraphics[scale=0.7]{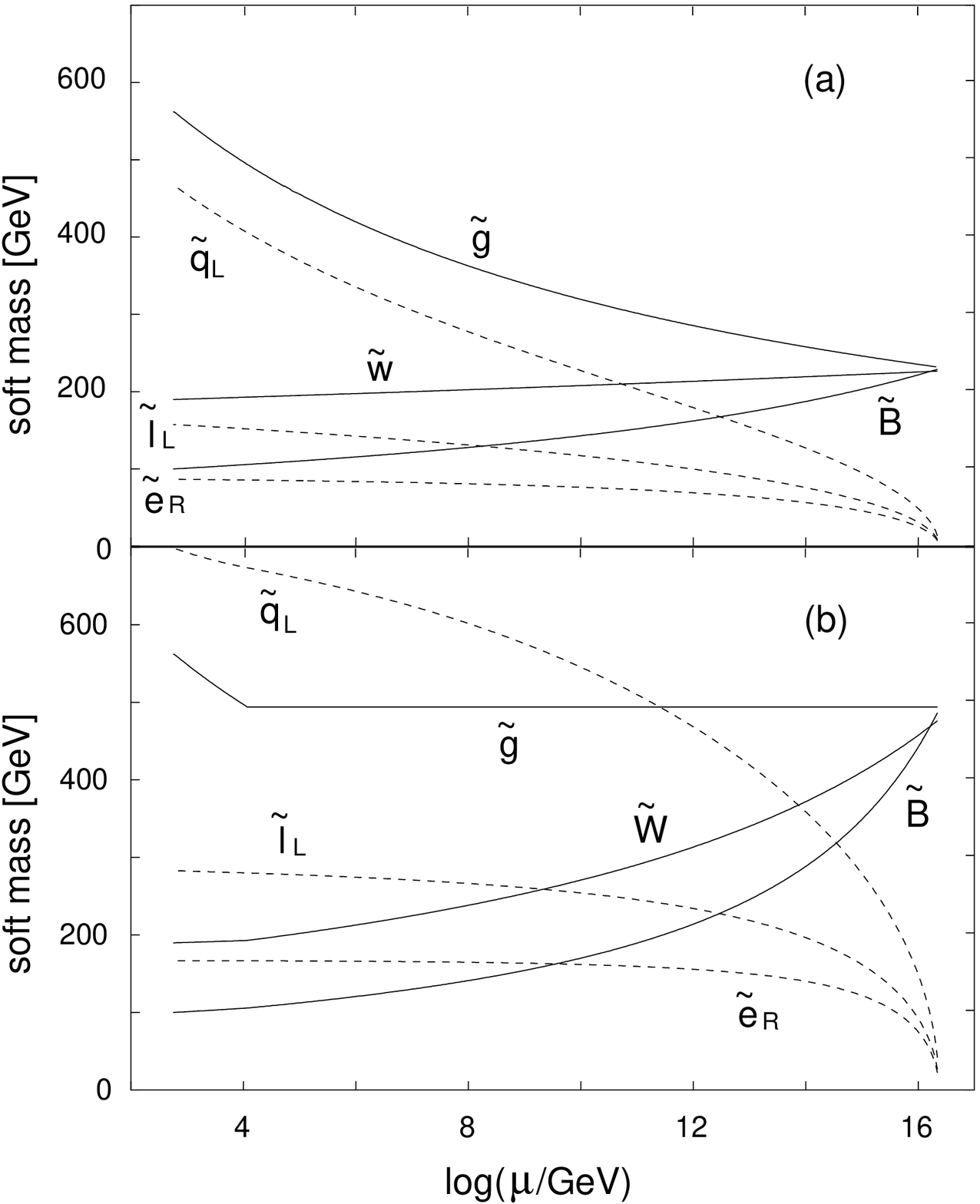}
   \end{center}
   \caption{RG evolutions of soft supersymmetry parameters where the B-ino 
     mass is fixed to be 100 GeV at the electroweak scale and scalar masses 
     are set to zero at the GUT scale.  Here the number of the extra matter 
     multiplets is $N_{\mbox{ex}} = 0$ (pure MSSM case) for (a) and 
     $N_{\mbox{ex}} = 3$ above the scale $M_{\mbox{ex}} = 10^4\ {\rm GeV}$ 
     for (b).  The solid lines are gaugino masses and the dashed ones are 
     scalar masses.}
   \label{fig:RGE}
  \begin{center}
    \includegraphics[scale=0.8]{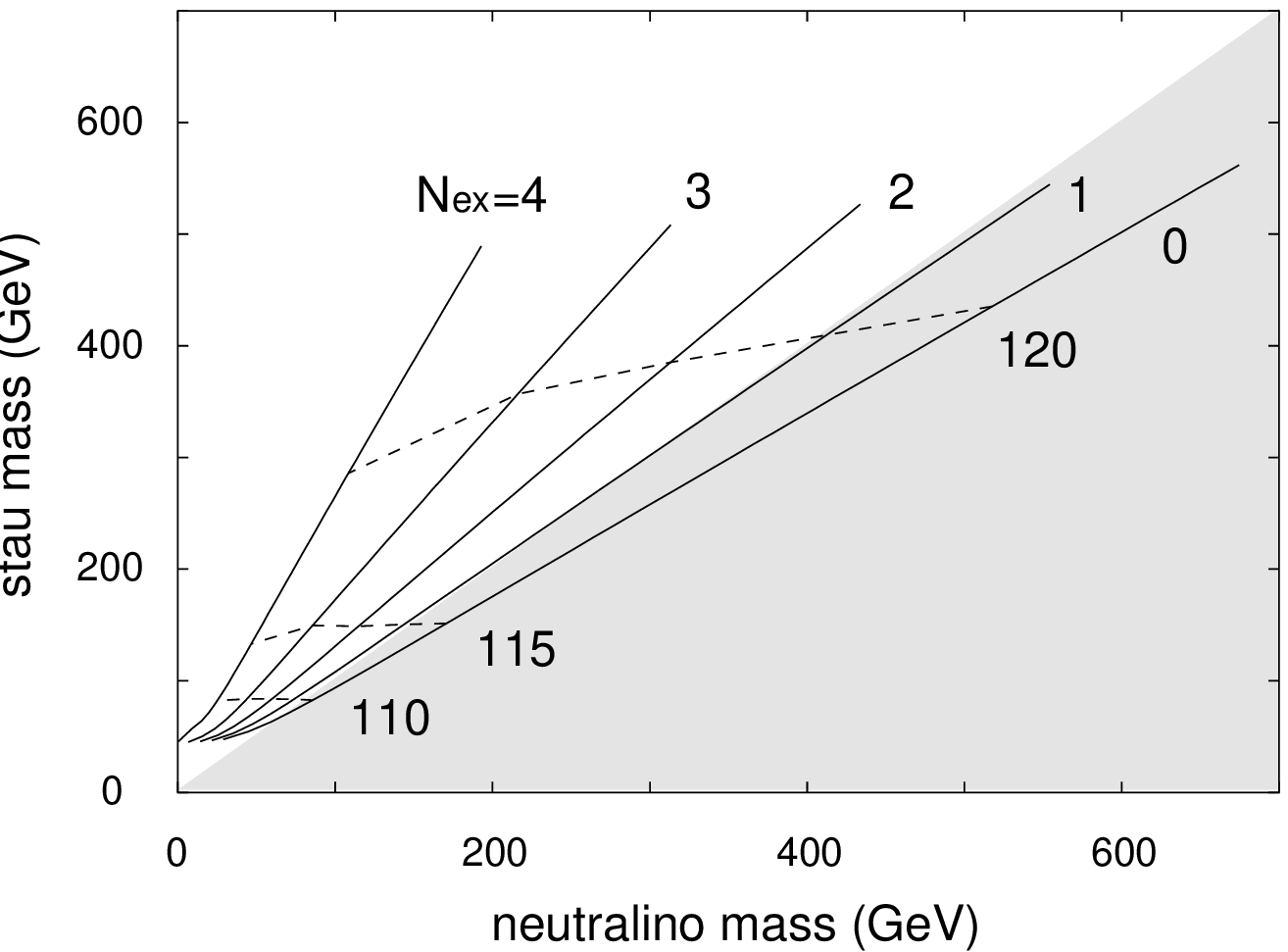}
  \end{center}
  \caption{The lightest neutralino mass v.s. the light stau mass at the 
    electroweak scale in no-scale model.  Each solid line corresponds to 
    a different number of the extra matter fields, $N_{\mbox{ex}} = 0-4$
    from right to left.  The mass scale of the extra matters is fixed at
    $M_{\mbox{ex}} = 10^4\ {\rm GeV}$, and the gaugigno mass at the GUT 
    scale is taken in the region of $M_{1/2} = 100-1500\ {\rm GeV}$.  
    Here we take $\tan\beta = 10$.  The contours of the Higgs mass are 
    also shown in the graph(dashed).  The shadow region is cosmologically 
    disfavored, in which the stau mass is lighter than the neutralino 
    mass.  
  }
  \label{fig:noscale}
\end{figure}

It is then straightforward to solve the RGEs of this system. Fig. \ref{fig:RGE}
demonstrate how the RG evolution changes in the presence of the extra
matter fields. Here the gaugino masses at the electroweak scale are
taken to be the same between the two specific cases of $N_{\mbox{ex}}$
($N_{\mbox{ex}} = 0$ for Fig. \ref{fig:RGE}(a) and $N_{\mbox{ex}} = 3$ for
(b)). To emphasize the effects of the RG evolution, we assume that the
soft scalar masses vanish at the boundary of the RG evolution, which
is assumed to be the GUT scale.  Also the decoupling scale of the
extra matter fields is chosen, somewhat arbitrarily, to $M_{\mbox{ex}}
= 10^{4}\ {\rm GeV}$ and $\tan\beta$, which is the ratio of the two
Higgs vacuum expectation values, is taken $\tan\beta = 10$. We also
assume that the gaugino masses have a common origin, that is,
universal gaugino mass. First, we find that the gaugino masses at the
low-energy are suppressed in the existence of the extra matter fields,
which is easily understood by noticing that the gauge couplings are
less asymptotic free at ultra-violet (UV) region. Put another way, in
order to obtain the same gaugino masses at the low-energy scale, one
has to start with a larger gaugino mass at the high-energy
scale. Combined with the fact that the gauge coupling constants are
also large in the UV side, the scalar fields acquire their soft masses
at the UV scale, which thus significantly enhance
the ratio of the sfermion masses with respect to the gaugino
masses  when compared with the case of no extra
matter.  

Here it is instructive to give analytic formulae for the soft
masses. For instance the $SU(2)_L$ gaugino mass and the soft scalar
masses of the first two generations are solved, when we impose a
universal soft scalar mass, $m_0$, and a universal gaugino mass, 
$M_{1/2}$ at the GUT scale $\sim 2 \times
10^{16}$ GeV, as:
\begin{eqnarray}
  M_2(\mbox{EW})&\simeq & 0.34 M_{1/2}, 
 \\
  m_{\tilde{q}}^2({\rm EW})
  &\simeq& m_0^2 +
  2.6 M_{1/2}^2 \simeq m_0^2 + 22 M_2^2(\mbox{EW}) ,
  \\
  m_{\tilde{\ell}_L}^2({\rm EW}) &\simeq& m_0^2 
  +
  0.35 M_{1/2}^2 \simeq m_0^2 + 3.0 M_2^2(\mbox{EW}) ,
 \\
  m_{\tilde{\ell}_R}^2({\rm EW}) &\simeq& m_0^2 
  +
  0.12 M_{1/2}^2 \simeq m_0^2 + 1.0 M_2^2(\mbox{EW}).
\end{eqnarray}
where we take $N_{\mbox{ex}} = 3$
and $M_{\mbox{ex}} = 10^4$ GeV.  The argument ``EW'' represents that 
they are quantities evaluated at the electroweak scale. In practice, we 
have set the renormalization scale at 500 GeV. 
These formulae should be compared with those of the MSSM case:
\begin{eqnarray}
  M_2(\mbox{EW})&\simeq & 0.84 M_{1/2}, 
 \\
  m_{\tilde{q}}^2({\rm EW})
  &\simeq& m_0^2 +
  4.9 M_{1/2}^2 \simeq m_0^2 + 6.9 M_2^2(\mbox{EW}) ,
  \\
  m_{\tilde{\ell}_L}^2({\rm EW}) &\simeq& m_0^2 
  +
  0.49 M_{1/2}^2 \simeq m_0^2+ 0.69 M_2^2(\mbox{EW}) ,
 \\
  m_{\tilde{\ell}_R}^2({\rm EW}) &\simeq& m_0^2 
  +
  0.15 M_{1/2}^2 \simeq m_0^2 + 0.21 M_2^2(\mbox{EW}).
\end{eqnarray}
The enhancement of the scalar masses relative to the gaugino masses at
the electroweak scale is apparent. In fact what happens here is that
the gaugino masses at low energy become smaller while the sfermion
masses do not change so much when $M_{1/2}$ is fixed, making the
scalar/gaugino mass ratio larger.

We should  note here that the contributions from
the extra matter fields become less significant for smaller $N_{\mbox{ex}}$
and for higher $M_{\mbox{ex}}$.  When $M_{\mbox{ex}}$ is larger than 
$10^{10}\ {\rm GeV}$, the effect becomes negligible.

\

Having established the increase of the sfermion masses, we now
consider phenomenological implications of the presence of the extra
matter multiplets. An immediate consequence is the modification of the
superparticle mass spectrum. This will be particularly important in
the future program to determine the mediation mechanism of the
supersymmetry breaking by tracing the RG flow to higher scale with the
superparticle masses which, we hope, will be measured at future
collider experiments as input parameters.  In this process, one has to
keep in mind that the presence of the extra matter fields can
drastically change the RG flow from that of the MSSM. We note that the
effect of the extra matter fields cannot be absorbed by the lift of the
universal scalar mass $m_0$, rather it gives a richer structure of the
superparticle mass spectrum.

In SUSY models, the experimental bound on the lightest CP-even Higgs
boson mass is known to provide a rather severe constraint on the soft
mass parameters, that is, larger soft masses are favored to enhance
the Higgs mass.  In fact, the Higgs mass bound from LEP II experiment
cannot be satisfied at tree level and radiative corrections play an
important role.  The radiative corrections mainly depend on the stop
masses and a larger stop mass yields a heavier Higgs mass~\cite{HiggsMass}.  
Thus the addition of the extra matter fields can
significantly relax the constraint on the parameter space from the
Higgs boson mass bound.  We will give an explicit example shortly.

Another implication we would like to discuss is on the so-called no-scale 
scenario. In this scenario, the soft masses satisfy the following 
no-scale boundary conditions: 
\begin{itemize}
\item vanishing soft scalar masses: $m_0=0$,
\item vanishing trilinear scalar couplings: $A=0$,
\item (generally) non-vanishing Higgs mixing parameter: $B$,
\item non-vanishing gaugino masses: $M_{1/2}$.
\end{itemize}
These are given at some fundamental scale, which we assume to be the GUT 
scale. 

With the MSSM matter contents ($N_{\mbox{ex}}=0$), the right-handed
slepton obtains a mass of $m^2_{\tilde{\ell_R}}(\mbox{EW}) \approx
0.84M_1^2(\mbox{EW})$, and thus it is smaller than
$M_1(\mbox{EW})$. Therefore the B-ino like neutralino can be lighter
than the right-handed slepton only when there is substantial mixing in
the neutralino mass matrix, which is the case when the gaugino mass is
not larger than the $Z$-mass scale.  In fact, a severe upperbound on 
the neutralino mass is obtained from the requirement that it becomes the
LSP, as was shown in~\cite{IKYY,KLNPY,Drees-Nojiri}.

In Fig. \ref{fig:noscale}, the masses of the lightest neutralino and 
the lighter stau at the electroweak scale are shown.  Here we take 
$N_{\mbox{ex}} = 0 - 4$, $\tan\beta = 10$ and 
$M_{\mbox{ex}} = 10^{4}\ {\rm GeV}$.  We also take the gaugino mass 
at the GUT scale in the range $M_{1/2} = 100 - 1500\ {\rm GeV}$.  
The comparison of the masses of the two
superparticles yields the region allowed by the cosmological argument
that the stable LSP should be neutral.  The shadow region is
excluded, as the charged stau is the LSP. One readily finds that in
the MSSM case ($N_{\mbox{ex}}=0$), the allowed region is very
restricted, where the upperbound of the neutralino mass is about 70
GeV. The region becomes somewhat enhanced for lower $\tan \beta$ ({\em
  e.g.} the upperbound becomes 110 GeV for $\tan \beta=3$), but still
the allowed region is quite limited.  On the other hand, as the number
of the extra matter increases, the allowed region where the lightest
neutralino (which is B-ino-like) becomes the LSP becomes drastically
larger. In fact, one finds that for $N_{\mbox{ex}} \gtrsim 2$ the
lightest neutralino always becomes the LSP.

The contour of the Higgs mass of $m_h = 115\ {\rm GeV}$, which roughly
corresponds to the present experimental lower
bound~\cite{HiggsMassBound}, is also drawn in the same figure.  Here
we have used {\tt FeynHiggsFast} package~\cite{FeynHiggsFast} to
compute the Higgs mass. We can see that the constraint from the Higgs
mass becomes relaxed significantly as $N_{\mbox{ex}}$ increases.  In
fact, in the MSSM case (i.e. with no extra matters), the
cosmologically allowed region does not satisfy the Higgs mass bound,
and thus the whole region is excluded.\footnote{This is also the case
  for lower $\tan\beta$. We checked this explicitly for $\tan
  \beta=3$.} However in the presence of the extra matter fields, the
sfermions, especially the stop, become heavier compared to the gaugino
masses, and thus the Higgs mass bound gives less restrictive
constraint on the gaugino-like neutralino mass $m_{\chi_1^0}$.

In Table \ref{table:quantities}, comparison of various quantities is 
made for $N_{\mbox{ex}}=0-4$ when $M_1$ at the electroweak scale is 
fixed to be 100 GeV.  Recall that the scalar masses increase as 
$N_{\mbox{ex}}$ increases.  Thus the SUSY contributions to low 
energy observables reduce for larger $N_{\mbox{ex}}$.  
In fact the branching ratio of $b \to s \gamma$ gradually approaches 
to the value of SM prediction as $N_{\mbox{ex}}$ becomes larger.  
Also the SUSY contribution to the muon $g-2$ is decreased.  
On the other hand, the Higgs mass becomes larger by the the enhancement 
of the radiative corrections and the constraint from the mass is relaxed, 
as was already mentioned.  

\
Some of the features discussed above are quantitatively modified when 
 $\tan\beta$ is large.  A crucial difference comes from the fact that the 
Yukawa coupling of the tau lepton is enhanced by $\tan\beta$, and becomes 
significantly large when $\tan\beta$ is large.  The large Yukawa coupling 
reduces the stau mass at low energy scale through the RG flow, and 
hence the requirement that the LSP should be neutral gives a stronger 
constraint on the parameter space.  We explicitly checked the case of 
$\tan\beta = 30$.  We found that the stau mass is reduced by about 100 GeV 
for $\tan\beta = 30$ while the neutralino mass is almost unchanged.  
As a result, $N_{\mbox{ex}} = 0, 1$ are 
completely excluded by cosmological argument.  For $N_{\mbox{ex}} = 2$, 
only the region where the lightest neutralino mass is heavier than 300 GeV 
is allowed.  The constraint is somewhat relaxed for $N_{\mbox{ex}} = 3$, 
with the neutralino mass required to be larger than 100 GeV.
Almost all regions are allowed for $N_{\mbox{ex}} = 4$.  
At the same time, the contour lines of the Higgs mass are also lowered 
about 100 GeV on the stau v.s. the neutralino mass line.

In the above analysis, we have implicitly taken the top mass $m_t$ to
be 175 GeV. We also analyzed the case of $m_t=180$ GeV. We found that
the lines of the stau-neutralino masses are almost intact because the
effects of the top Yukawa couplings come through the determination of
the supersymmetric higgsino mass parameter, $\mu$, and in our case
$\mu$ is large and thus its effects are decoupled. On the other hand,
the Higgs mass changes significantly since the radiative corrections are
proportional to the 4-th power of the top mass. In fact the computed
Higgs mass is found to increase by about 2 or 3 GeV.

Finally we would like to make a brief comment on how the recent WMAP
result \cite{Bennett:2003bz,Spergel:2003cb} on the abundance of the
dark matter, $\Omega_{DM} h^2 \approx 0.11$, affects on our
scenario. For the bino-like LSP, its relic abundance calculated under
the standard thermal history of the Universe tends to be larger than
the WMAP result. One way to evade this difficulty is invoke efficient
coannihilation \cite{Mizuta-Yamaguchi,Falk} with the sleptons. It
requires that the stau mass is quite degenerate with the neutralino
mass. In our case, this is achieved by appropriately adjusting the
mass $M_{\mbox{ex}}$ to make the effect of the extra matter fields
less significant. Another possibility is to assume non-standard
thermal history of the Universe below the weak scale, such as
late-time entropy production to dilute the abundance of the
neutralinos. Note that when the LSP were stau, the entropy production
would not be able to reduce its relic density enough to survive the
severe constraint from charged massive stable particle searches
\cite{kudo}.  We leave further study on the issue of the relic
abundance of the neutralino LSP for future publication.

\

To summarize, we have pointed out that the addition of the extra vector-like
chiral multiplets can significantly change the RG evolution of the soft SUSY
breaking masses in the MSSM. In particular, we found that the sfermion masses 
are enhanced relative to the gaugino masses at low energy. We also illustrated
phenomenological implications of this effect, such as the change of the
superparticle mass spectrum, the enhancement of the lightest CP-even
Higgs mass and also some impact on the no-scale scenario.

\section*{Acknowledgment}
This work was supported in part by the Grants-in-aid from the Ministry
of Education, Culture, Sports, Science and Technology, Japan, No.12047201 and
No.14046201.
ME thanks the Japan Society for the Promotion of Science for financial 
support.

\section*{Note Added}
After submitting the paper, we learnt that the addition of the extra
matter fields was also considered in Ref. \cite{Jack:2003sx} in a different
context, {\it i.e.} in the comparison with higher order effects to the
RG evolution. We thank D.R.T. Jones for drawing our attention to this paper. 


\begin{table}[htbp]
\begin{center}
\begin{tabular}{c|ccc}
{$N_{\mbox{ex}}$}
& {$Br(b \to s \gamma)\ (\times 10^{-4})$}
& {$a_\mu = (g_\mu-2)/2\ (\times 10^{-9})$}
& {$m_h^0$} \\
\hline
{$0$}  & {$3.0$}  & {$4.3$}  & {$111$} \\
{$1$}  & {$3.0$}  & {$3.4$}  & {$113$} \\
{$2$}  & {$3.0$}  & {$2.7$}  & {$114$} \\
{$3$}  & {$3.1$}  & {$1.7$}  & {$116$} \\
{$4$}  & {$3.2$}  & {$0.8$}  & {$119$} \\
\end{tabular}
\caption{Various quantities for $N_{\mbox{ex}} = 0-4$ with $M_1(\mbox{EW})$ 
fixed to be 100 GeV.}
\label{table:quantities}
\end{center}
\end{table}

\end{document}